\documentclass[11pt,twoside]{article}
\newcommand{\affil}[1]{$^{\rm #1}$}
\usepackage[authoryear]{natbib}
\usepackage[usenames]{color}
\bibpunct{(}{)}{;}{a}{}{,}
\usepackage{graphicx}
\date{} 
\title{\large\bf\flushleft Reality of Moving Groups in the Galaxy and Chemically 
Tagging the Galactic Disk }
\author{\parbox{\textwidth}{\flushleft
\vspace{-0.5cm}
{\it G.M. De Silva\affil{1}, K.C. Freeman\affil{2}, J. Bland-Hawthorn\affil{3}, and M. Asplund\affil{4}}\\
\vspace{0.4cm}
{\small \affil{1}\,European Southern Observatory, Karl-Schwarzschild Str 2. D-85748 Garching, Germany.\\ Email:  gdesilva@eso.org}\\
{\small \affil{2}\,Research School of Astronomy and Astrophysics, Mount Stromlo Observatory, Australian National University. ACT 2611. Australia.}\\
{\small \affil{3}\,Institute of Astronomy, University of Sydney, NSW 2006, Australia. }\\
{\small \affil{4}\,MPA, Karl-Schwarzschild Str 1., D-85748 Garching, Germany }
}}
\begin{document}
%
\begin{changemargin}{.8cm}{.5cm}
\begin{minipage}{.9\textwidth}
\vspace{-1cm}
\maketitle
%
%
\small{{\bf Abstract.} The existence of old dispersed stellar groups within the Milky Way disk is still controversial. Are they the debris of ancient star-forming aggregates, or short-lived artifacts of dynamical origin? With detailed elemental abundance measurements from high quality spectroscopic data, we show that at least one such old dispersed stellar group is a true relic of an earlier phase of star formation. The identification of other such relic structures will provide essential information for probing the evolutionary history of the Milky Way disk.
}

\medskip
\medskip
\end{minipage}
\end{changemargin}
\small


\section{Chemical Tagging}   
A major goal of near-field cosmology is to determine the physical sequence of events involved in the formation of the Galactic disk. Unlike for the stellar halo, the disk's dissipative history means most of its dynamical information is lost. Any dynamical probing of the disk can only provide information up until the epoch of last dissipation or dynamical scattering event. However the chemical information survives. If star-forming aggregates have unique \emph{chemical signatures}, we can use the method of chemical tagging to tag disk stars to a common formation event (Freeman \& Bland-Hawthorn 2002). Detailed chemical abundance patterns offer the possibility to reconstruct now dispersed stellar aggregates of the protogalactic disk and so improve our basic understanding of the disk formation process (Bland-Hawthorn \& Freeman 2004).

\bigskip

The first test for the chemical tagging technique is that galactic open clusters are chemically homogeneous. Our studies on the Hyades open cluster and the very old open cluster Collinder 261 showed they are highly chemically homogenous with little or no intrinsic star-to-star scatter for a range of key elements (De Silva et al.~2006, 2007). This implies that pollution is negligible and that the stars' natal chemistry is preserved. Therefore stellar chemical abundance patterns can be regarded as {\it fossil signatures} of the history of the disk's stellar content. The next step for the viability of chemical tagging is to establish chemical homogeneity in unbound groups, such as superclusters and moving groups, which retain some kinematical identity. The final step is to chemically identify groups which have no dynamical identity. 

\section{Moving groups and superclusters}

The concept of moving groups and superclusters was introduced by Eggen in the 1960s. As a star cluster orbits around the Galaxy, it disperses into a tube-like structure around the Galaxy plane, and eventually dissolve into the background. The tube-like unbound groups of stars occupying extended regions of the Galaxy were defined by  Eggen as superclusters. If the Sun happens to be inside this tube,  the group members will appear to us all over the sky, but identifiable as a group through their common space velocities. Such groups were defined as a moving group, which is a subset of the larger supercluster. These dispersed groups are therefore the in-between step from bound clusters to field stars. 

\bigskip

One such dispersed star cluster is the moving group associated with the star HR1614 (Eggen 1998, Feltzing \& Holmberg 2000). However, doubt has been expressed about the reality of Eggen's moving groups and superclusters (e.g.~Taylor 2000), suggesting that these kinematically defined structures arise due to disk dynamical effects, such as spiral density waves. If so, these groups would be transient objects, consisting of a mixture of field stars. Our recent study on the HR1614 moving group stars show that this group is chemically homogeneous and distinguishable from the field stars (De Silva et al.~2007).  The high level of chemical homogeneity supports the case that the HR1614 moving group is a true remnants of a dispersing star cluster.

\bigskip

Not all co-moving stellar groups in the disk are remnants of clusters. For example the kinematically defined Hercules stream cannot be distinguished from the field stars in elemental abundance space (Bensby et al.~2007),  implying that the Hercules stream has a dynamical origin where random field stars have been swept into a common orbit. It is not a remnant of a star-forming event. Conversely, the HR1614 moving group stars are chemically homogeneous and chemically distinguishable from the field abundance patterns, indicating that it is a dispersed relic of an earlier star-forming event. When inspected in velocity space alone, both the HR1614 moving group and the Hercules stream satisfy the requirements for a dispersing cluster remnant, while detailed abundance results show they have very different origins. These examples demonstrate that chemical information is essential to identify dispersed cluster members, and that dynamical information alone cannot be used to uncover the true history behind any stellar stream in the Galactic disk.





\end{document}